\documentstyle [12pt] {article}
\begin{document}

\def\ri{\rightarrow}
\def\L{{\cal L}}
\def\tr{{\rm Tr}}
\def\Q{{\cal Q}}
\def\non{\nonumber}
\def\kl{K_L\to \pi^+\pi^-\gamma}
\def\ks{K_S\to\pi^+\pi^-\gamma}
\def\kp{K^+\to\pi^+\pi^0\gamma}
\def\eps{\epsilon'_{+-\gamma}}
\def\im{ {\rm Im}}
\def\re{{\rm Re}}
\def\pr{{\sl Phys. Rev.}~}
\def\prl{{\sl Phys. Rev. Lett.}~}
\def\pl{{\sl Phys. Lett.}~}
\def\np{{\sl Nucl. Phys.}~}
\def\zp{{\sl Z. Phys.}~}

\font\el=cmbx10 scaled \magstep2
{\obeylines
\hfill ITP-SB-93-32
\hfill IP-ASTP-20-93
\hfill May, 1993}

\vskip 2.0 cm

\centerline {{\el Direct CP Violation in $K_L\to\pi^+\pi^-\gamma$}}

\medskip
\bigskip
\medskip

\centerline{\bf Hai-Yang Cheng}

\medskip
\centerline{ Institute of Physics, Academia Sinica}
\centerline{Taipei, Taiwan 11529, Republic of China}

\centerline{and}

\medskip
\centerline{ Institute for Theoretical Physics, State University of
New York}
\centerline{Stony Brook, New York 11794, USA}

\bigskip
\bigskip

\centerline{\bf Abstract}
\bigskip

    CP violation due to interference between $K_L$ and $K_S$ decays into $\pi
^+\pi^-\gamma$ is analyzed in the Standard Model.
The CP-violating parameter $\epsilon'_{+-
\gamma}$, which is the difference between $\eta_{+-\gamma}$ and $\eta_{+-}$,
receives dominant contributions from $K^0-\bar{K}^0$ mixing and the gluon
penguin diagram; its magnitude is calculated to be $10^{-2}\epsilon$ at a
typical region of the Dalitz plot. Genuine direct CP violation in $K_L\to\pi^+
\pi^-\gamma$
decay originating from the electromagnetic penguin diagram is of order
$10^{-4}\epsilon$.

\pagebreak

   Recently, interference between $K_L$ and $K_S$ into the final state $\pi^+
\pi^-\gamma$ has been observed for the first time in experiment 731 at
Fermilab [1]. A new CP-violating parameter $\eta_{+-\gamma}$ to be defined
later is extracted from this experiment to be $(2.15\pm 0.26\pm 0.20)\times
10^{-3}$ in magnitude and $(72\pm 23\pm 17)^\circ$ in phase angle. Writing
$\eta_{+-\gamma}=\eta_{+-}+\epsilon'_{+-\gamma}$
and using the known value of $\eta_{+-}$ measured from
$K_L\to\pi^+\pi^-$ decay, a limit on the parameter $\epsilon'_{+-\gamma}$
in the decay mode $\kl$ is obtained: $|\eps|/\epsilon<0.3\,$. The purpose of
this Letter is to study the theoretical expectation of  $\eps$ in the
Standard Model.

   The radiative decays $K\to \pi\pi\gamma$ receive two distinct
contributions: inner bremsstrahlung (IB) and direct emission (DE). Under
Lorentz and gauge invariance, the general expression for the decay $K(k)\to
\pi(p_1)\pi(p_2)\gamma(q,\epsilon)$ reads
\begin{eqnarray}
A(K\to\pi\pi\gamma)=\,A_{\rm IB}+A_{\rm DE},
\end{eqnarray}
where
\begin{eqnarray}
A_{\rm IB}={\cal A}e\left({p_1\cdot \epsilon\over p_1\cdot q}-{p_2\cdot
\epsilon\over p_2\cdot q}\right),
\end{eqnarray}
\begin{eqnarray}
A_{\rm DE}={\cal B}[ie\epsilon_{\mu\nu\rho\sigma}p^\mu_1 p^\nu_2q^\rho
\epsilon^\sigma]+{\cal C}e[(p_1\cdot\epsilon)(p_2\cdot q)-(p_2\cdot\epsilon)
(p_1\cdot q)].
\end{eqnarray}
The first term in $A_{\rm DE}$ is caused by a pure magnetic
transition, while the second term corresponds to an electric transition. The
coefficients ${\cal A,~B}$ and ${\cal C}$ are in general momentum dependent.
In the leading multipole expansion, ${\cal B}$ corresponds to a magnetic dipole
(M1) transition, whereas ${\cal A,~C}$ to electric dipole (E1) transitions.
{}From chiral perturbation theory (ChPT) point of view, DE
amplitudes are of most interest because they provide an ideal place to test
higher-order weak chiral Lagrangians. In ChPT, the DE radiative decay $K\to
\pi\pi\gamma$ cannot be generated from the lowest order $p^2$ chiral Lagrangian
because Lorentz and gauge invariance requires at least three powers of
momenta in the amplitude.

  The most general $p^4$ CP-invariant $\Delta S=1$ non-anomalous electroweak
chiral Lagrangian with one external photon field which satisfies
the constraints of chiral and $CPS$ symmetry
\footnote{$CPS$ is a discrete symmetry which is the product of ordinary $CP$
with a switching symmetry $S$, which switches the $d$- and $s$-quark fields;
see Ref.[2].}
has the expression [3]
\begin{eqnarray}
\L^{\Delta S=1}_{\rm non-anom}=i\left( {2\over f_\pi^2}\right)g_8eF^{\mu\nu}
[\omega_1\tr(\lambda_6 L_\mu L_\nu Q)+\omega_2\tr(\lambda_6 L_\nu Q L_\mu)]
\end{eqnarray}
for normal intrinsic parity transitions, while anomalous Lagrangian terms
for the odd intrinsic parity sector are [4]
\begin{eqnarray}
\L^{\Delta S=1}_{\rm anom} &=& ia\left( {2\over f_\pi^2}\right)g_8e\tilde{F}^{
\mu\nu}\tr(QL_\mu)\tr(\lambda_6 L_\nu)  \non \\
&+& ib\left( {2\over f_\pi^2}\right)g_8e\tilde{F}^{
\mu\nu}\tr(QU^\dagger L_\mu U)\tr(\lambda_6 L_\nu)  \non \\
&+& ic\left( {2\over f_\pi^2}\right)g_8e\tilde{F}^{
\mu\nu}\tr\left(\lambda_6[UQU^\dagger,~L_\mu L_\nu]\right),
\end{eqnarray}
where $\tilde{F}_{\mu\nu}\equiv \epsilon_{\mu\nu\alpha\beta}F^{\alpha\beta}$,
$Q={\rm diag}({2\over 3},-{1\over 3},-{1\over 3})$, $L_\mu\equiv (D_\mu
U)U^\dagger$ with $D_\mu U=\partial_\mu U-ieA_\mu[Q,~U]$ is an $SU(3)_R$
singlet, and
\begin{eqnarray}
&& U = \exp\left(2i{\phi\over f_\pi}\right),~~~~f_\pi=132\,{\rm MeV}, \non \\
&& \phi\equiv{1\over\sqrt{2}}\phi^a\lambda^a=\left(\matrix{ {\pi^0\over\sqrt
{2}}+{\eta\over\sqrt{6}} & \pi^+ & K^+  \cr \pi^- & -{\pi^0\over\sqrt{2}}+
{\eta\over \sqrt{6}} & K^0  \cr K^- & \bar{K}^0 & -\sqrt{2\over 3}\eta  \cr}
\right).
\end{eqnarray}
In Eqs.(4) and (5), $g_8$ is the octet weak coupling constant appearing in
the lowest order CP-invariant $\Delta S=1$ weak chiral Lagrangian
\begin{eqnarray}
\L^{(2)}_W=-g_8\tr(\lambda_6L_\mu L^\mu).
\end{eqnarray}
Presently, there is only one experimental information on the couping constants
appearing in $\L^{\Delta S=1}_{\rm non-anom}$.
{}From the BNL measurement of the $K^+\to\pi^+ e^+e^-$ decay rate [5],
one finds a scale-independent result
\begin{eqnarray}
\omega_1+2\omega_2-12L_9\simeq -7.5\times 10^{-3},
\end{eqnarray}
where $L_9$ is one of the coupling constants in the $p^4$ strong-interaction
chiral Lagrangian coupled to the external vector and axial-vector gauge fields
[6]. When combining with the empirical value of $L_9=6.7\times 10^{-3}$ at the
renormalization scale $\mu=m_\rho$ [6], this leads to
\begin{eqnarray}
\omega^r_1(\mu=m_\rho)+2\omega_2^r(\mu=m_\rho)\simeq 0.074\,,
\end{eqnarray}
where the superscript $r$ means renormalized coupling constant.

  In principle, the unknown coupling constants $\omega_1,~\omega_2,~a,~b$ and
$c$ should be determined from various low-energy hadronic
processes. However, in the limit of large $N_c$ ($N_c$ being the number of
quark color degrees of freedom), these couplings become theoretically
manageable at least to the zeroth order of $\alpha_s$ [4]. This is based on
bosonization, factorization, and the $\Delta S=1$ effective weak Hamiltonian
at the quark level. It is found that in the large-$N_c$ approach
\footnote{A different large-$N_c$ prediction $\omega_1=\omega_2=8L_9$ is
obtained in Ref.[8].}
\begin{eqnarray}
\omega_1=\omega_2=4L_9={N_c\over 12\pi^2},
\end{eqnarray}
and
\begin{eqnarray}
a=2b=4c={N_c\over 12\pi^2}.
\end{eqnarray}
The result (11) first obtained in Ref.[4] was recently confirmed by Ref.[7].
Several remarks are in order: (i) The couplings $a,~b$ and $c$ are determined
by chiral anomalies and hence are free of gluonic corrections.
(ii) It was pointed out in Ref.[3] that the relation
$\omega_2=4L_9$ must hold at least for the divergent parts of the counterterm
coupling constants because they must render the divergent loop amplitudes
finite. Evidently, this relation is preserved in the large-$N_c$ approach.
(iii) The large-$N_c$ prediction $\omega+2\omega_2=0.076$ is in
remarkable agreement with (9), though the latter is renormalization scale
dependent. The scale-indepenent prediction $\omega_1+2\omega_2-12L_9=0$
is also in good agreement with (8) in view of the fact that $12L_9(\mu=m_\rho)
\simeq 0.08$.

   At the $p^4$ level, there are two different contributions to direct
emission of $K\to\pi\pi\gamma$: contact-term contributions induced by
$\L^{\Delta S=1}_{\rm non-anom}$ and $\L^{\Delta S=1}_{\rm anom}$, and the
long-distance pole contributions generated by the weak $K-\pi$ transition
and the anomalous $\pi\pi\pi\gamma$ interaction
governed by the anomalous Wess-Zumino-Witten term [9]
\begin{eqnarray}
\L_{\rm WZW}=-{N_c\over 3\pi^2f_\pi^3}e\epsilon^{\mu\nu\alpha\beta}A_\mu\tr(
Q\partial_\nu \phi\partial_\alpha\phi\partial_\beta\phi)+\cdots.
\end{eqnarray}
We now focus on the decays $K_{L,S}\to \pi^+\pi^-\gamma$. For reasons to be
explained later, the $\omega_1$ term in $\L^{\Delta S=1}_{\rm non-anom}$ does
not contribute to any radiative decays of $K_{L,S}$ with a real photon
emission. In the limit of CP symmetry, the $\omega_2$ term does not contribute
 to $\kl$, whereas $\L^{\Delta S=1}_{\rm anom}$ makes no contribution to
$K_S\to\pi^+\pi^-\gamma$ (see Ref.[4] for details). We thus see that if CP is
good, $\kl$ is a pure M1 transition,
while $K_S\to\pi^+\pi^-\gamma$ is caused by electric transitions.
Unfortunately, one cannot make a definite prediction for the direct emission
of $\kl$ despite the fact that it is originated from chiral anomalies. This is
because the $\pi^0$ and $\eta$ pole contributions cancel each other due to the
Gell-Mann-Okubo mass relation $m^2_\eta={1\over 3}(4m^2_K-m^2_\pi)$, a
phenomenon also known to the decay $K_L\to\gamma\gamma$ (for a review, see
Ref.[10]). Therefore, contrary to the common belief,
\footnote{The DE amplitude of $\kl$ is widely assumed to be dominated by
pseudoscalar meson poles in the literature; needlessly to say, this
assumption has no ground in ChPT.}
the direct emission of $\kl$ at the $p^4$ level is solely determined by the
anomalous contact term $\L^{\Delta S=1}_{\rm anom}$
and has a branching ratio
\footnote{This number is obtained from Eqs.(4.21) and (4.11) of Ref.[4]. Since
we are working in the leading order in $1/N_c$ expansion, chiral loops can be
neglected.}
of $2\times 10^{-4}$,
which is too large by an order of magnitude when compared
with the recent experimental value of $(3.19\pm 0.16)\times 10^{-5}$ measured
in E731 experiment at Fermilab [11]. This implies that a {\it large} and {\it
destructive} $p^6$
contribution is called for in order to accommodate data. At order ${\cal O}
(p^6)$, several effects should be taken into consideration: SU(3)
breaking, $\eta-\eta'$ mixing,
\footnote{The SU(3)-singlet $\eta'$ is integrated out in
ChPT, but its effect on $\kl$ will manifest in the coefficients of higher
order Lagrangian terms. In practice, the pole amplitude is evaluated at the
$p^4$ level by explicitly incorporationg the $\eta'$ pole. Theoretical
uncertainties include SU(3) breaking in the decay constants $f_\eta,~f_{\eta'}
$ and in the matrix element $<\eta|\L_W|K>$ [12], $\eta-\eta'$ mixing angle,
and nonet-symmetry breaking in the matrix element $<\eta'|\L_W|K>$ [13].}
momentum dependence of vertices, and higher order anomalous Lagrangian
$\L^{\Delta S=1}_{\rm anom}$.
In the absence of knowledge of the $p^6$ Lagrangians, no reliable
prediction on the direct emission of $\kl$ can be made at present.

   In the presence of CP violation, $K_L$ is allowed to have an E1 transition
into the final state $\pi^+\pi^-\gamma$. Possible CP-violating effects in $\pi
\pi\gamma$ decay of neutral kaons have been studied by many authors [14,15].
Some effects such as asymmetry between $\pi^+$ and $\pi^-$ spectra, photon
polarization, require going beyond the dipole approximation and are expected
to be small. At the leading dipole level, CP nonconservation can be
observed through the $K_L-K_S$ interference [15]. It is natural to consider the
CP-violating parameter
\begin{eqnarray}
\eta_{+-\gamma}=\,{A(\kl,E1)\over A(K_S\to\pi^+\pi^-\gamma,E1)}.
\end{eqnarray}
Note that M1 transitions do not play a role here.
Since $\ks$ decay is predominated by IB, as will be shown later, and since
IB amplitudes have the expressions
\begin{eqnarray}
A(\ks)_{\rm IB} &=& -eA(K_S\to\pi^+\pi^-)\left({p_+\cdot \epsilon\over p_+
\cdot q}-{p_-\cdot \epsilon\over p_-\cdot q}\right),  \non \\
A(\kl)_{\rm IB} &=& -eA(K_L\to\pi^+\pi^-)\left({p_+\cdot \epsilon\over p_+
\cdot q}-{p_-\cdot \epsilon\over p_-\cdot q}\right),
\end{eqnarray}
we are led to
\begin{eqnarray}
\eta_{+-\gamma}=\eta_{+-}+\epsilon'_{+-\gamma}=\eta_{+-}+{A(\kl,E1)_{\rm DE}
\over A(\ks,E1)_{\rm IB}},
\end{eqnarray}
where $\eta_{+-}=A(K_L\to\pi^+\pi^-)/A(K_S\to\pi^+\pi^-)=\epsilon+\epsilon'$
characterizes CP noninvariance in $K_L\to\pi^+\pi^-$ decay.
Because in the convention of ChPT, $|K_{1,2}>={1\over\sqrt{2}}(|K^0>\mp|\bar{K}
^0>)$, we write the wave functions of $K_{L,S}$ as
\begin{eqnarray}
|K_L> &=& {(1+\bar{\epsilon})|K^0>+(1-\bar{\epsilon})|\bar{K}^0>\over\sqrt{2(1
+|\bar{\epsilon}|^2)}}={|K_2>+\bar{\epsilon}\,|K_1>\over \sqrt{1+|\bar{
\epsilon}|^2}}, \non  \\
|K_S> &=& {(1+\bar{\epsilon})|K^0>-(1-\bar{\epsilon})|\bar{K}^0>\over\sqrt{2(1
+|\bar{\epsilon}|^2)}}={|K_1>+\bar{\epsilon}\,|K_2>\over \sqrt{1+|\bar{
\epsilon}|^2}},
\end{eqnarray}
where the parameter $\bar{\epsilon}$ (though not physical) measures the amount
of CP
violation in the neutral kaon wave functions. It follows from Eq.(15) that
\begin{eqnarray}
\epsilon'_{+-\gamma}=\,{A(K_1\to\pi^+\pi^-\gamma,E1)_{\rm DE}\over A(\ks,E1)_{
\rm IB}}\left(\bar{\epsilon}+{A(K_2\to\pi^+\pi^-\gamma,E1)_{\rm DE}\over A(
K_1\to\pi^+\pi^-\gamma,E1)_{\rm DE}}\right).
\end{eqnarray}
As $\eps$ does receive contributions from CP nonconservation in $K^0-\bar
{K}^0$ mixing, it is not a pure ``direct'' CP-violating effect. We will come
back to this point later.

   Even at this stage, we can make a crude estimate on the magnitude of $\eps$
simply based on the experimental observation of the decay rates of IB and DE
in $K_L$ and/or $K^+$. First of all, the two terms in the parentheses of
Eq.(17) are of order $\epsilon$. Second, the DE rate is naively expected
to be suppressed relative
to that of IB. However, IB of $\kl$ is subject to CP violation as can be seen
from Eq.(14). Likewise, the IB amplitude of $K^+\to\pi^+\pi^0\gamma$ is a
$\Delta I={3\over 2}$
transition and is thus suppressed by a factor of $A(K_S\to \pi\pi(I=0))/A(K_S
\to\pi\pi(I=2))\cong 22$. Consequently, the DE rates in the $\pi\pi\gamma$
decays of $K_L$ and $K^+$ are not very small compared to that of IB, and they
even dominate in the $K_L$ case. This explains why
structure-dependent DE effects can be practically seen in the radiative
decays of $K_L$ and $K^\pm$. Experimentally [11,16],
\begin{eqnarray}
Br(\kl)_{\rm IB} &=&(1.49\pm 0.08)\times 10^{-5}, \non \\
Br(\kl)_{\rm DE} &=&(3.19\pm 0.16)\times 10^{-5},  \non \\
Br(\kp)_{\rm tot} &=& (2.75\pm 0.15)\times 10^{-4},  \\
Br(\kp)_{\rm DE} &=& (1.8\pm 0.4)\times 10^{-5}. \non
\end{eqnarray}
The decay $\kp$ has a larger IB rate because the $\Delta I={3\over 2}$
suppression is less severe than the corresponding CP-violation one in the
counterpart of $\kl$. For $\ks$ decay, its IB amplitude is no longer subject
to CP violation or $\Delta I={3\over 2}$ suppression. As a consequence, this
decay mode is totally dominated by the IB mechanism. Although the ratio of
$A(K_1\to\pi^+\pi^-\gamma,E1)_{\rm DE}/ A(\ks,E1)_{\rm IB}$ depends on the
region of the Dalitz plot, we see from Eq.(18) that it is roughly of order
$10^{-2}$ after taking into account the suppression factor for IB amplitudes
in $K_L$ and $K^\pm$ decays. We thus have $\eps\sim 10^{-2}\epsilon$,
which is in accordance with more accurate calculations presented below.

   In the Standard Model, there exist two different contributions of amplitude
CP violation to the DE amplitude of $K_2\to\pi^+\pi^-\gamma$: one from the
gluon-penguin diagram, and the other from the loop $s\to d\gamma$ diagram
(so-called electromagnetic penguin diagram); that is,
\begin{eqnarray}
A(K_2\to\pi^+\pi^-\gamma,E1)_{\rm DE}=A(K_2\to\pi^+\pi^-\gamma,E1)_{\rm DE}^{
\rm gluon}+A(K_2\to\pi^+\pi^-\gamma,E1)_{\rm DE}^{\rm em}.
\end{eqnarray}
We first focus on the gluon-penguin diagram. The lowest order CP-odd $\Delta
S=1$ weak Lagrangian reads
\begin{eqnarray}
\L_W^-=\,-ig_8'\tr(\lambda_7L_\mu L^\mu).
\end{eqnarray}
The Gell-Mann matrix $\lambda_7$ ensures that the decay $K_2\to\pi\pi$ is
allowed by
$\L_W^-$, but not $K_1\to\pi\pi$. Unlike the CP-even octet
coupling constant $g_8$, the CP-odd one $g_8'$ presumably only receives
contributions from the short-distance QCD penguin
diagram. The CP-odd $\Delta S=1$ effective
Hamiltonian induced by the gluon penguin diagram has the structure
\footnote{To the first order in chiral expansion, the penguin operator $Q_6$
is equivalent to $-4(Q_2-Q_1)v^2/\Lambda^2_{\chi}$ [17], where $v$
characterizes the quark order parameter $<\bar{q}q>$, and $\Lambda_\chi\sim 1$
GeV is a chiral-symmetry breaking scale.}
\begin{eqnarray}
{\cal H}^-_{\rm peng}=i{G_F\over\sqrt{2}}\sum_{i=u,c,t}{\rm Im}(V^*_{is}V_{id})
I_i(Q_2-Q_1-h.c.),
\end{eqnarray}
where $Q_1=(\bar{s}d)(\bar{u}u),~Q_2=(\bar{s}u)(\bar{u}d)$, with $(\bar{q}_i
q_j)\equiv\bar{q}_i\gamma_\mu(1-\gamma_5)q_j$. Since the bosonization of the
quark current $(\bar{q}_iq_j)$ is ${i\over 2}f^2_\pi(L_\mu)_{ji}$ (see
Eq.(3.13) of Ref.[4] for higher order corrections), it follows that
the leading-order chiral representation of $(Q_2-Q_1-h.c.)$ is equivalent to
$(f_\pi^2/4\pi)\tr(\lambda_7 L_\mu L^\mu)$. The coupling constant $g_8'$ is
related to ${G_F\over\sqrt{2}}\sum\im(V^*_{is}V_{id})I_i$, but it
turns out that we do not have to know the exact expression of
the penguin coefficients $I_i$. The argument goes as follows.
Starting from Eq.(21) and following the
prescription in Ref.[4] for the derivation of large-$N_c$ $p^4$
electroweak chiral Lagrangians, it is easily seen that the
relevant CP-odd $p^4$ Lagrangians have the same structure as the CP-even ones
$\L^{\Delta S=1}_{\rm non-anom}$ and $\L^{\Delta S=1}_{\rm anom}$ given before
except for the replacement of $g_8$ by $ig_8'$ and the Gell-Mann matrix
$\lambda_6$ by $\lambda_7$. For example, the CP-odd $p^4$ non-anomalous terms
can be read directly from Eq.(4) as
\begin{eqnarray}
\L^-_{\rm non-anom}=-\left( {2\over f_\pi^2}\right)g_8'eF^{\mu\nu}
[\omega_1\tr(\lambda_7 L_\mu L_\nu Q)+\omega_2\tr(\lambda_7 L_\nu Q L_\mu)].
\end{eqnarray}
It follows from Eqs.(4) and (22) that
\begin{eqnarray}
{A(K_2\to\pi^+\pi^-\gamma)_{\rm DE}^{\rm gluon}\over A(K_1\to\pi^+\pi^-\gamma)
_{\rm DE}}=\,i{g_8'\over g_8}.
\end{eqnarray}
However, from Eqs.(7) and (20) we obtain
\begin{eqnarray}
{A(K_2\to\pi\pi(I=0))\over A(K_1\to\pi\pi(I=0))}=\,i{g_8'\over g_8}.
\end{eqnarray}
This together with (23) leads to
\begin{eqnarray}
{A(K_2\to\pi^+\pi^-\gamma)_{\rm DE}^{\rm gluon}\over A(K_1\to\pi^+\pi^-\gamma)
_{\rm DE}}={A(K_2\to\pi\pi(I=0))\over A(K_1\to\pi\pi(I=0))}={A_0-A_0^*\over
A_0+A_0^*}=i{{\rm Im}A_0\over {\rm Re}A_0},
\end{eqnarray}
where $A_0\equiv A(K^0\to\pi\pi(I=0))$ and use of $CP|\bar{K}^0>=-|K^0>$
has been made.
This, when combining with $\bar{\epsilon}$, gives rise to precisely the
well-known CP-violating parameter $\epsilon$ [18]
\begin{eqnarray}
\epsilon\cong \bar{\epsilon}+i{{\rm Im}A_0\over {\rm Re}A_0}.
\end{eqnarray}
This result has the desired feature: Neither $\bar{\epsilon}$ nor $i{\rm Im}
A_0/{\rm Re}A_0$ is rephasing invariant, but their combination is a physical
quantity. For example, consider a phase redefinition of the $s$ qaurk field,
$s\to se^{i\alpha}$. It follows $|K^0>\to |K^0>e^{-i\alpha},~|\bar{K}^0>\to |
\bar{K}^0>e^{i\alpha}$. Recasting $\epsilon$ into the form [18]
\begin{eqnarray}
\epsilon={1\over\sqrt{2}}e^{i\theta}\left( {1\over 2}{ {\rm Im}M_{12}\over
{\rm Re}M_{12}}+{ {\rm Im}A_0\over\re A_0}\right),~~~\theta=\tan^{-1}{2\Delta
M\over\Gamma_S}\simeq 45^\circ,
\end{eqnarray}
with $M_{12}=<K^0|{\cal H}^{\Delta S=2}_{\rm eff}|\bar{K}^0>$ being the neutral
kaon mass matrix element,
we see that $\epsilon$ is rephasing invariant as $\im M_{12}/\re M_{12}\to(\im
M_{12}/\re M_{12}+2\alpha)$ and $\im A_0/\re A_0\to (\im A_0/\re A_0-\alpha)$
under infinitesimal phase transformation.

   Up to now, we have
\begin{eqnarray}
\epsilon'_{+-\gamma}=\,{A(K_1\to\pi^+\pi^-\gamma,E1)_{\rm DE}\over A(\ks,E1)_{
\rm IB}}\left(\epsilon+{A(K_2\to\pi^+\pi^-\gamma,E1)_{\rm DE}^{\rm em}\over A(
K_1\to\pi^+\pi^-\gamma,E1)_{\rm DE}}\right).
\end{eqnarray}
Genuine direct CP violation (i.e. amplitude CP violation) comes from the
second term in $\eps$ originating from the loop $s\to d\gamma$ transition. The
corresponding Lagrangian is given by [19]
\begin{eqnarray}
\L^{\rm EM} = {G_F\over\sqrt{2}}\,{e\over 16\pi^2}\sum_{i=u,c,t}V^*_{is}
V_{id}\left\{ c_7(x_i)Q_7+F(x_i)Q_T\right\}+h.c.,
\end{eqnarray}
where $x_i=m^2_i/M_W^2$, and
\begin{eqnarray}
 Q_7 &=& \bar{s}\gamma_\mu(1-\gamma_5)d\,\partial_\nu F^{\mu\nu}, \non \\
 Q_T &=& i[m_s\bar{s}\sigma_{\mu\nu}(1-\gamma_5)d+m_d\bar{s}
\sigma_{\mu\nu}(1+\gamma_5)d]F^{\mu\nu}, \\
 F(x) &=& {(8x^2+5x-7)x\over 12(x-1)^3}-{(3x-2)x^2\over 2(x-1)^4}\ln x. \non
\end{eqnarray}
Since the chiral realization of $\bar{s}\gamma_\mu(1-\gamma_5)d$ is
proportional to $(L_\mu)_{23}$, it is straightforward to show that the chiral
representation of $\bar{s}\gamma_\mu(1-\gamma_5)d\partial_\nu F^{\mu\nu}$
is exactly the $\omega_1$ term in Eq.(4) [3]. Of course, besides the
short-distance $s\to d\gamma$ effect,
the coefficient $\omega_1$ also receives long-distance contributions.
However, the fact that
$\partial_\mu F^{\mu\nu}=0$ for a real photon emission suffices to lead us to
conclude that, as was promised before,  the $\omega_1$ term
$\tr(\lambda_6 L_\mu L_\nu Q)F^{\mu\nu}$ does not contribute to any
radiative decays of neutral kaons with one on-shell photon, as can be
explicitly checked.

   The tensor operator $Q_T$-induced DE amplitude of $\kl$ is given by
\begin{eqnarray}
A(\kl)_{\rm DE}^{\rm em}=\,iG_F\,{e\over 16\pi^2}\im(V^*_{ts}V_{td})F(x_t)<
\pi^+\pi^-\gamma|Q_T|K^0>,
\end{eqnarray}
where $F(x_i)$ is obviously dominated by the top quark. Note
that the matrix element $<\pi^+\pi^-|\bar{s}\sigma_{\mu\nu}\gamma_5d|K^0>$ is
related to $<\pi^+\pi^-|\bar{s}\sigma_{\mu\nu}d|K^0>$ through the identity
\begin{eqnarray}
\sigma_{\mu\nu}\gamma_5=-{i\over 2}\epsilon_{\mu\nu\alpha\beta}\sigma^{\alpha
\beta}.
\end{eqnarray}
Therefore, the general expression for the matrix element of the tensor operator
$Q_T$ reads
\begin{eqnarray}
<\pi^+\pi^-\gamma|Q_T|K^0> &\cong& 2m_s\epsilon^\mu q^\nu<\pi^+\pi^-|\bar{s}
\sigma_{\mu\nu}(1-\gamma_5)d|K^0>   \non \\
&=& \beta[(p_+\cdot\epsilon)(p_-\cdot q)-(p_-\cdot\epsilon)(p_+\cdot q)-i
\epsilon_{\mu\nu\rho\sigma}p^\mu_+p^\nu_-q^\rho\epsilon^\sigma].
\end{eqnarray}
Our task is to evaluate the coefficient $\beta$. It is interesting to note
that the $K^0\to\pi\pi$ amplitude induced by the operator $\tilde{Q}_T=[
\bar{s}\sigma_{\mu\nu}(1-\gamma_5)d+h.c.]$ also occurs in the Weinberg
model of CP violation in which three Higgs doublets are introduced to
generate CP nonconservation via Higgs exchange [20]. Previously, this
amplitude is evaluated through the use of current algebra and the bag-model
calculation of the $K-\pi$ transition, but it is subject to
large theoretical uncertainties (for a review, see Ref.[21]). Here we
will follow Ref.[22]
to use the chiral language to get an estimate of the coefficient $\beta$.

   The lowest order chiral realization of $\tilde{Q}_T$ is $\tr(\lambda
_6U)$. One may be tempted to think that $\beta$ is thus fixed in
this way. However, this is not the case. The reason is being that an
additional tadpole diagram arising from the strong  $K\pi
\to K\pi$ scattering followed by a $K\to$ vacuum weak transition should be
included [23,21], and it exactly compensates the direct $K^0\to\pi\pi$
amplitude induced by $\tilde{Q}_T$. The vanishing $\tilde{Q}_T$-generated
$K^0\to\pi^+\pi^-\gamma$ amplitude to the lowest order in chiral symmetry
also can be understood from the chiral point
of view of $Q_T$. As pointed out in Ref.[22], in order to write down its
chiral representation one needs at least three $U$ matrices to get two
derivatives to match with $F_{\mu\nu}$. Hence, the CP-violating $\kl$
decay induced by the $s\to d\gamma$ mechanism is necessary of higher order
chiral effect.

   The chiral structure of $O_T$ is not the same as $\L^{\Delta S=1}_{\rm
non-anom}$ and $\L^{\Delta S=1}_{\rm anom}$ since it vanishes in the chiral
limit.
Among many possible realizations, a typical one for E1 transitions is [22]
\footnote{The charge matrix $Q$ given in Ref.[22] is diag$(1,0,0)$. We use this
chance to point out that a crucial factor of $(1/4\pi)$ is likely missing in
Eq.(15) of Ref.[22]. Consequently, CP-violating asymmetry for $K^\pm\to
\pi^\pm\pi^0$ and $K^\pm\to\pi^\pm\pi^0\gamma$ is overestimated there by a
factor of $(4\pi)^2$.}
\begin{eqnarray}
O_T\to{\cal O}_T\sim \gamma F^{\mu\nu}\bigg( M\partial_\mu U[Q,~U^\dagger]
\partial_
\nu U+\partial_\mu U^\dagger[Q,~U]\partial_\nu U^\dagger M\bigg)_{23},
\end{eqnarray}
with $M={\rm diag}(m_u,m_d,m_s)$. There are of course other chiral terms for
M1 transitions, but as noted in passing, the M1 amplitude is related to the E1
one through Eq.(33). The coefficient $\gamma$ is estimated by naive dimensional
analysis to be $f_\pi$.
\footnote{There are several slightly different versions of naive dimensional
analysis; we use the updated one [24]. Note that the sign of the coefficient
$\gamma$ or $\beta$ is not fixed by this method.}
For our purpose of estimation, it suffices to apply (34) to fix the parameter
$\beta$. We find
\begin{eqnarray}
<\pi^+\pi^-\gamma|{\cal O}_T|K^0>_{\rm DE}\simeq -{8\over f_\pi^3}\,\gamma m_s[
(p_+\cdot\epsilon)(p_-\cdot q)-(p_-\cdot\epsilon)(p_+\cdot q)],
\end{eqnarray}
and hence
\begin{eqnarray}
\beta\sim -{8\over f_\pi^2}\,m_s.
\end{eqnarray}
Note that the same result is obtained by working out the higher order
realization of the operator $\tilde{Q}_T$. In the Chau-Keung parametrization
of the quark-mixing matrix [25],
\begin{eqnarray}
\im(V_{ts}^*V_{td})\simeq s_{13}s_{23}\sin\delta_{13},
\end{eqnarray}
where we have adopted the notation of the Particle Data Group [16]. Putting
everything together, we get
\begin{eqnarray}
A(\kl)^{\rm em}_{\rm DE}=-i{G_Fm_s\over 2\pi^2f_\pi^2}(s_{13}s_{23}\sin\delta
_{13})F(x_t)e[(p_+\cdot\epsilon)(p_-\cdot q)-(p_-\cdot\epsilon)(p_+\cdot q)].
\end{eqnarray}
As for the DE amplitude of $K_1\to\pi^+\pi^-\gamma$, we obtain from Eqs.(4)
and (10) that (see also Eq.(4.21) of Ref.[4])
\begin{eqnarray}
A(K_1\to\pi^+\pi^-\gamma)_{\rm DE}=-{8g_8\over \sqrt{2}\pi^2f_\pi^5}\,e[(p_+
\cdot\epsilon)(p_-\cdot q)-(p_-\cdot\epsilon)(p_+\cdot q)].
\end{eqnarray}
Numerically, we thus have
\footnote{This ratio is rephasing invariant, though $\im(V_{ts}^*V_{td})$ by
itself is not.}
\begin{eqnarray}
{A(\kl,E1)^{\rm em}_{\rm DE}\over A(K_1\to\pi^+\pi^-\gamma,E1)_{\rm DE}}=-i
4.3\times 10^{-5}\sin\delta_{13},
\end{eqnarray}
where uses have been made of $m_s=175$ MeV, $s_{23}=0.044,~s_{13}/s_{23}=0.1,
{}~F(x_t)=0.34$ for $m_t=150$ GeV, and the experimental value [26]
\begin{eqnarray}
g_8=-0.26\times 10^{-8}m^2_K.
\end{eqnarray}

   Finally, we need to know the ratio of $A(K_1\to\pi^+\pi^-\gamma,E1)_{\rm
DE}/A(K_S\to\pi^+\pi^-\gamma,E1)_{\rm IB}$, which is obtained from Eqs.(14)
and (39) to be
\begin{eqnarray}
{A(K_1\to\pi^+\pi^-\gamma,E1)_{\rm DE}\over A(K_S\to\pi^+\pi^-\gamma,E1)_{\rm
IB}}={4\sqrt{2}g_8\over \pi^2f_\pi^5A(K_S\to\pi^+\pi^-)}\,(p_+\cdot q)(p_-\cdot
q).
\end{eqnarray}
In principle, this ratio and hence $\eps$ is maximized at the region of the
Dalitz plot where the photon has its highest energy. However, we instead choose
a typical Dalitz point, say $E_\gamma=90$ MeV, to get a more realistic estimate
of $\eps$. A simple calculation gives
\begin{eqnarray}
(p_+\cdot q)(p_-\cdot q)=\,4.5\times 10^{-4}~{\rm GeV}^4~~~{\rm at}~E_\gamma=
90\,{\rm MeV}.
\end{eqnarray}
This together with the experimental value [26]
\begin{eqnarray}
A(K_S\to \pi^+\pi^-)=3.88\times 10^{-7}~{\rm GeV},
\end{eqnarray}
and Eqs.(42), (43) yields
\begin{eqnarray}
\left.{A(K_1\to\pi^+\pi^-\gamma,E1)_{\rm DE}\over A(K_S\to\pi^+\pi^-\gamma,E1)
_{\rm IB}}\right|_{E_\gamma=90\,{\rm MeV}}=-1.1\times 10^{-2}.
\end{eqnarray}
By virture of Eqs.(28), (40) and (45), we eventually arrive at the result
\begin{eqnarray}
\eps(E_\gamma=90\,{\rm MeV})=-1.1\times 10^{-2}(\epsilon-i4.3\times 10^{-5}
\sin\delta_{13}).
\end{eqnarray}
Note that an estimate of $\eps\sim 10^{-2}\epsilon$ is also reached in
Ref.[12] but with a different reasoning.

    We conclude that $\eps$ receives dominant contributions from  $K^0-\bar
{K}^0$ mixing and the gluon penguin diagram, and its magnitude is estimated
 to be $10^{-2}\epsilon$ at a
typical Dalitz point. Genuine direct CP violation in $K_L\to\pi^+\pi^-\gamma$
decay originating from the electromagnetic penguin diagram is of order
$10^{-4}\epsilon$. This is close to the expected lower bound
of the direct CP-violating parameter $\epsilon'$ in the two pion decays of
neutral kaons in the Standard Model.

\pagebreak

\noindent {\bf Acknowledgments}:~~~
The author wishes to thank Prof. C. N. Yang and the
Institute for Theoretical Physics at Stony Brook for their hospitality
during his stay there for sabbatical leave.
This research was supported in part by the
National Science Council of ROC under Contract No.  NSC82-0208-M001-001Y.

\vskip 2.0cm

\centerline{\bf REFERENCES}

\bigskip

\begin{enumerate}

\item E.J. Ramberg {\it et al.}, \prl {\bf 70}, 2529 (1993).

\item C. Bernard, T. Draper, A. Soni, H.D. Politzer, and M. Wise, \pr {\bf
D32}, 2343 (1985).

\item G. Ecker, A. Pich, and E. de Rafael, \np {\bf B291}, 692 (1987); \pl
{\bf B189}, 363 (1987); \np {\bf B303}, 665 (1988).

\item H.Y. Cheng, \pr {\bf D42}, 72 (1990).

\item C. Alliegro {\it et al.,} \prl {\bf 68}, 278 (1992).

\item J. Gasser and H. Leutwyler, \np {\bf B250}, 465 (1985).

\item J. Bijnens, G. Ecker, and A. Pich, \pl {\bf B286}, 341 (1992).

\item C. Bruno and J. Prades, \zp {\bf C57}, 585 (1993).

\item See, e.g., N.K. Pak and P. Rossi, \np {\bf B250}, 279 (1985).

\item L. Littenberg and G. Valencia, Fermilab-Pub-93/004-T (1993), to appear
in {\sl Ann. Rev. Nucl. Part. Sci.} volume 43.

\item E.J. Ramberg {\it et al.}, \prl {\bf 70}, 2525 (1993).

\item Y.C.R. Lin and G. Valencia, \pr {\bf D33}, 143 (1988).

\item H.Y. Cheng, \pl {\bf B245}, 122 (1990).

\item T.D. Lee and C.S. Wu, {\sl Ann. Rev. Nucl. Part. Sci.} {\bf 16}, 511
(1967).

\item G. Costa and P.K. Kabir, {\sl IL Nuovo Cimento}, {\bf 51A}, 564 (1967);
L. Sehgal and L. Wolfenstein, \pr {\bf 162}, 1362 (1967).

\item Particle Data Group, \pr {\bf D45}, S1 (1992).

\item S. Fajfer and J.-M. G\'erard, \zp {\bf C42}, 425 (1989);
H.Y. Cheng, \pr {\bf D43}, 1579 (1991).

\item L.L. Chau, {\sl Phys. Rep.} {\bf 95}, 1 (1983).

\item T. Inami and C.S. Lim, {\sl Prog. Theor. Phys.} {\bf 65}, 297 (1981).

\item S. Weinberg, \prl {\bf 37}, 657 (1976).

\item H.Y. Cheng, {\sl Int. J. Mod. Phys.} {\bf A7}, 1059 (1992).

\item C.O. Dib and R.D. Peccei, \pl {\bf B249}, 325 (1990).

\item J.F. Donoghue and B.R. Holstein, \pr {\bf D32}, 1152 (1985).

\item H. Georgi, \pl {\bf B298}, 187 (1993).

\item L.L. Chau and W.Y. Keung, \prl {\bf 53}, 1802 (1984).

\item H.Y. Cheng, {\sl Int. J. Mod. Phys.} {\bf A4}, 495 (1989).

\end{enumerate}

\end{document}